\documentclass[conference]{IEEEtran}
\usepackage[colorinlistoftodos]{todonotes}
\usepackage{multirow}

\usepackage{subfigure}

\usepackage{cite}

\usepackage{amsmath}
\usepackage{amssymb}
\usepackage[ruled,linesnumbered]{algorithm2e}

\begin{document}
%
\title{Principled Multilayer Network Embedding}

\author{
\IEEEauthorblockN{Weiyi Liu\IEEEauthorrefmark{1}\IEEEauthorrefmark{2}, 
Pin-Yu Chen\IEEEauthorrefmark{2},
Sailung Yeung\IEEEauthorrefmark{3},
Toyotaro Suzumura\IEEEauthorrefmark{2} and 
Lingli Chen\IEEEauthorrefmark{1}} \IEEEauthorblockA{
\IEEEauthorrefmark{1}University of Electronic Science and Technology of China\\}
\IEEEauthorblockA{\IEEEauthorrefmark{2}IBM Watson Research Center\\}
\IEEEauthorblockA{\IEEEauthorrefmark{3} Boston University\\}
Emails: weiyiliu@us.ibm.com, pin-yu.chen@ibm.com, yeungsl@bu.edu, suzumura@acm.org, lingli324@std.uestc.edu.cn}

\maketitle

\begin{abstract}
Multilayer network analysis has become a vital tool for understanding different relationships and their interactions in a complex system, where each layer in a multilayer network depicts the topological structure of a group of nodes corresponding to a particular relationship. The interactions among different layers imply how the interplay of different relations on the topology of each layer.
For a single-layer network, network embedding methods have been proposed to project the nodes in a network into a continuous vector space with a relatively small number of dimensions, where the space embeds the social representations among nodes. These algorithms have been proved to have a better performance on a variety of regular graph analysis tasks, such as link prediction, or multi-label classification.

In this paper, by extending a standard graph mining into multilayer network, we have proposed three methods (``network aggregation,'' ``results aggregation'' and ``layer co-analysis'') to project a multilayer network into a continuous vector space.
On one hand, without leveraging interactions among layers, ``network aggregation'' and ``results aggregation'' apply the standard network embedding method on the merged graph or each layer to find a vector space for multilayer network.
On the other hand, in order to consider the influence of interactions among layers, ``layer co-analysis'' expands any single-layer network embedding method to a multilayer network. By introducing the link transition probability based on information distance, this method not only uses the first and second order random walk to traverse on a layer, but also has the ability to traverse between layers by leveraging interactions.
From the evaluation, we have proved that comparing with regular link prediction methods, ``layer co-analysis'' achieved the best performance on most of the datasets, while ``network aggregation'' and ``results aggregation'' also have better performance than regular link prediction methods.

\end{abstract}


%
\IEEEpeerreviewmaketitle

\section{Introduction}\label{sec:introduction}

With the rise of the big data phenomenon (and in particular, the rise of social networking), graph mining has become necessary to analyze the diverse relationships between objects and data, and to understand the complex structures of the underlying graph.
Analyzing such complex systems is crucial in understanding how a social network forms, or in making predictions about future network behavior.

However, graph mining is sensitive to the topological structures of a network. For example, in social network analysis, graph mining can leverage topology information to identify communities in a network, but does not have the ability to determine if the topology contains noise (or other errors). One way to avoid this is using multilayer network \cite{boccaletti2014structure}. 
Multilayer network is a group of networks which depict multiple relationships between nodes, where each layer in the group represents a particular type of relationship \cite{de2013mathematical,boccaletti2014structure,loe2015comparison}. 

\begin{figure}[!t]
\centering
\includegraphics[width=\linewidth]{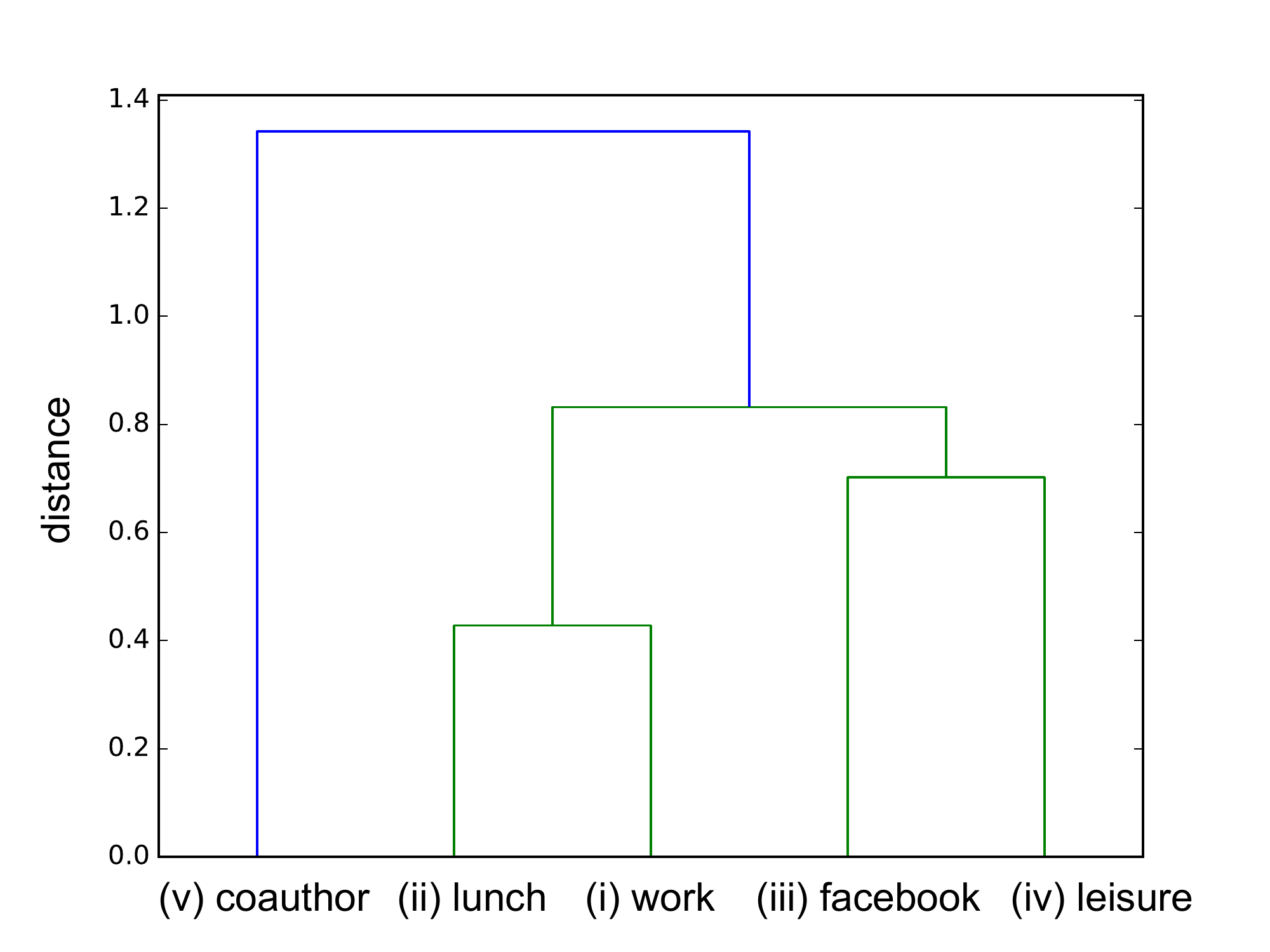}
\caption{Dendrogram of layers for the AUCS dataset.}
\label{intro-layers}
\end{figure}

Take the AUCS dataset \cite{kim2015community} as an example. The multiple layers represent five different relationship types between 61 employees of a university department: (i) coworking, (ii) having
lunch together, (iii) facebook friendship, (iv) spending leisure time together, and (v) coauthorship. Figure \ref{intro-layers} shows the distances among these different layers, calculated by \cite{de2015structural}, where the distance between layers is the distance value of the root node of the smallest subtree containing both layers. From Figure \ref{intro-layers}, we observe a number of key interactions: for example we observe that being coworkers and having lunch together (representing professional interactions) is strongly related through the low layer distance, and spending lesiure time together is likewise close to being connected on facebook (representing social interactions).
In general, by considering multiple relationship types, we argue that multilayer network inherently reflects essential interactions between nodes in a manner that is robust to the noise presents in any individual relationship type.

Nowadays, graph mining methods on multilayer network typically concentrate on different network granularities \cite{de2017community}, depending on the task. For example, on node/edge granularity, \cite{heaney2014multiplex} and \cite{hu2014conditions} focus on developing suitable centrality measures \cite{costenbader2003stability} like cross-layer degree centrality for multilayer network \cite{brodka2011degree,brodka2012analysis}. Brodka et.al \cite{kazienko2010individual,brodka2010method} proposed multilayered local clustering coefficient (MLCC) and cross-layer clustering coefficient (CLCC) to depict cluster coefficient \cite{zhou2005maximal} of a node in a multilayer network.
In contrast, the cluster level is often used for community detection \cite{boccaletti2014structure,de2015identifying,loe2015comparison,domenico2015identifying,chen2016multilayer,jeub2017local,deford2017spectral}, and the layer level used to analyze the interactions between different layer types \cite{de2015structural,benson2016higher}.

In this paper, we made an attempt to propose three novel graph mining methods for multilayer network by combining analysis at all levels of granularity that is suitable for a range of tasks. We achieve this via embedding the nodes of the multilayer network into a vector space. This vector space can be interpreted as a hidden metric space \cite{boguna2009navigability} for the multilayer network that naturally defines concepts of similarity between nodes, and captures many of the aspects of the original multilayer network, facilitating vector-based representation to a variety of machine learning algorithms to solve a range of tasks.

In standard network analysis, many such network embedding methods have been developed \cite{chen2017fast}, such as DeepWalk \cite{perozzi2014deepwalk}, LINE \cite{tang2015line} and node2vec \cite{grover2016node2vec}. These methods are all based upon generating samples of random walks over an input graph, with the properties of the random walk varying by method.
However, these methods are built on top of a single graph, and to the best of our knowledge, the graph embedding method for multilayer graph has not been rigorously explored.
Hence, we propose a generic multilayer graph embedding framework, which applies to any graph embedding method developed for single-layer graphs. In particular, we introduce three principled methods (``Network Aggregation,'' ``Results Aggregation'' and ``Network Co-analysis'') for extending graph embedding to multilayer networks:
\begin{itemize}
\item \textbf{Network Aggregation}: The assumption for this method is all edges from different layers are equal \cite{boccaletti2014structure,loe2015comparison}. Based on this assumption, this method aggregates all networks into a single (weighted) network (where multiple edges between nodes are not allowed) and then applies existing graph embedding algorithms to analysis the merged graph. Note that this merged graph no longer distinguishes between relationship types, as many edges for node pairs that share multiple edges in the multilayer network are not retained. 

\item \textbf{Results Aggregation}: The assumption for this method is different layers have totally different kinds of edges \cite{berlingerio2013abacus}, which implies that even if two nodes have edges in different layers, these edges are totally different and cannot be merged. Based on this assumption, this method applies graph embedding to each layer separately, then merges the vector spaces together to define a vector space of the multilayer network.

\item \textbf{Layer Co-analysis}: Unlike the first two methods which only consider inter-layer edges (network aggregation) or intra-layer edges (results aggregation), this method leverages interactions among different layers to allow for traversal between layers, and to retain the structure of each individual layer. Specifically, we introduce a new random walk method that is capable of traversing layers, thereby encoding important interactions between nodes and layers. The methodology of random walk in a multilayer network is a general extension of single-layer embedding methods. For example, node2vec and DeepWalk are both state-of-the-art single-layer graph embedding methods. As node2vec added to DeepWalk the ability to control the homophily and structural equivalence properties of random walk samples, we enable the ability for random walk samples to traverse multiple layers. In particular, we control the degree to which this occurs through the addition of the $r\in[0,1]$ parameter, which determines the probability that the next step of a random walk will traverse to a different layer in the multilayer network.
\end{itemize}

In Section \ref{method}, we gives detailed information on three approaches for performing embedding on multilayer networks. In Section \ref{eva}, we use experimental results to demonstrate that our methods for multilayer network embedding can improve upon the results of node2vec for a link prediction task. In Section \ref{sec:related}, we outline related work in single network embedding, and conclude in Section \ref{sec:conclusion} with discussion and future work. 


\section{Methods} \label{method}
\begin{figure}[!t]
\centering
\subfigure[Architecture of Network Aggregation]{
	\label{intro-NA}
    \includegraphics[width=\linewidth]{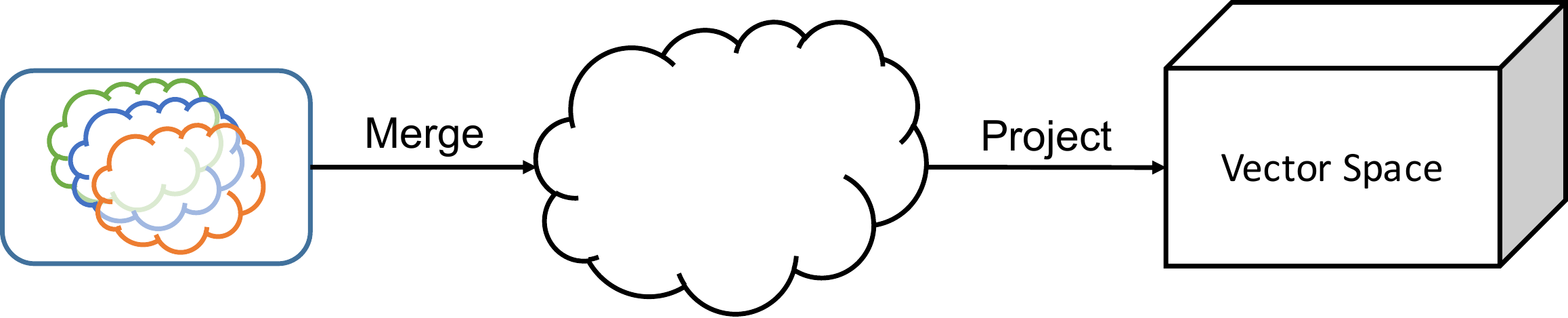}
}
\subfigure[Architecture of Results Aggregation]{
	\label{intro-RA}
    \includegraphics[width=\linewidth]{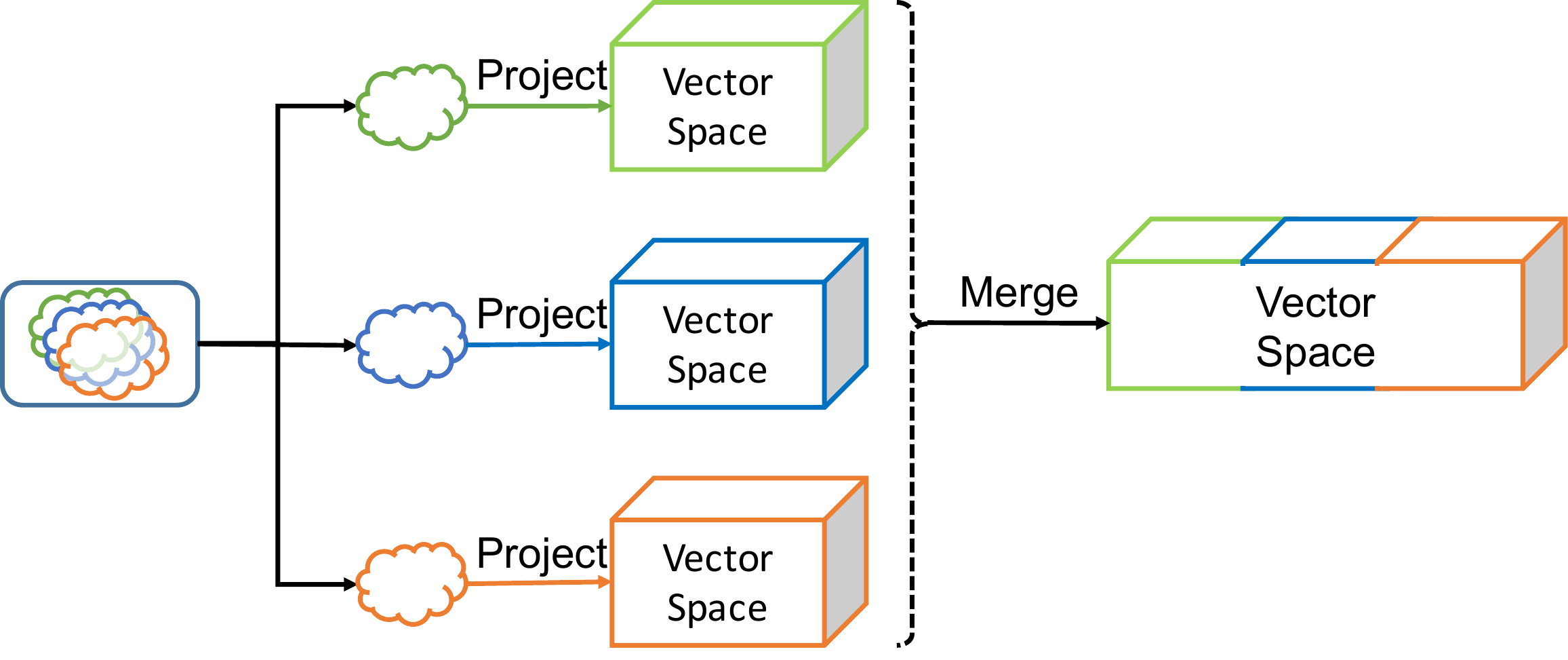}
}
\subfigure[Architecture of Layer Co-analysis]{
	\label{intro-LC}
    \includegraphics[width=\linewidth]{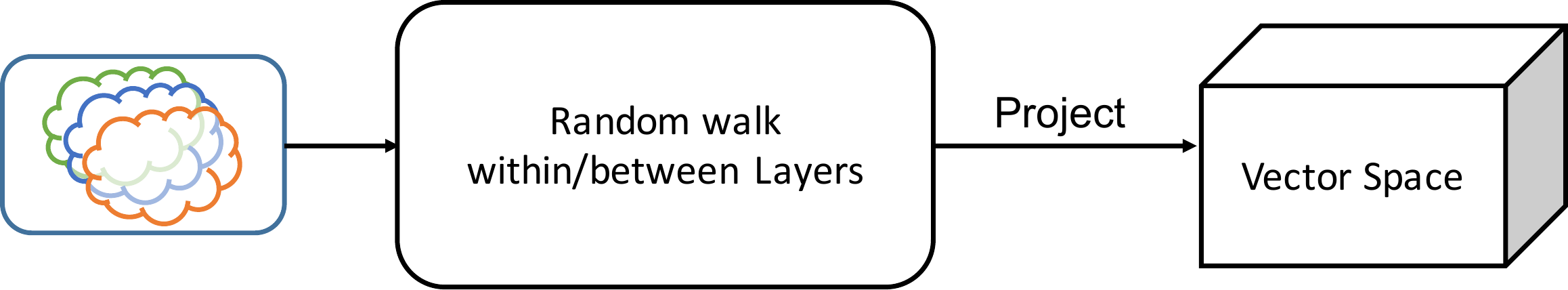}
}
\caption{The architecture of three methods
to do multilayer networks embedding.}
\label{med-overview}
\end{figure}

Given a multilayer network $MN=(V,L,A)$ with vertex set $V$, layer set $L$, and multilayer edge set $A$ (where $A\subseteq \left\{ (x,y,l)|x,y\in V,l\in L \right\}$ and we denote the existence of an edge between $x$ and $y$ in layer $l$ by $a^l_{xy}=1$), our task in vertex embedding is to learn a mapping function $f:V \to \mathbb{R}^d$ where $d$ is the chosen dimension of the vector space. We are effectively looking to find a function $f$ that represents the features of a vertex from the multilayer network. Figure \ref{med-overview} shows the three proposed architectures for projecting a multilayer network into a vector space. The following subsections provide details on the implementation and structure of each architecture, as well as the corresponding strategy for constructing the function $f$.

\subsection{Network Aggregation}\label{subsec:NA}
Network aggregation is the baseline of the proposed methods for multilayer network embedding, where the layers of the multilayer network are merged to obtain a single network, and the regular node2vec method is applied to the merged graph. Algorithm \ref{Alg:NA} represents the network aggregation process.

In this method, we have a network aggregation function $g: MN\to G$, i.e. a function that takes a multilayer MN and outputs a merged network $G=(V,E)$ that shares the vertex set $V$ with the multilayer network, but has an edge set $E=\left\{ { (i,j)|\sum _{ l\in L }^{  }{ a^{ l }_{ ij }\geq 1 }  } \right\} $ that disallows multi-edges. The mapping function $f$ is defined by training on the merged network $G$ using node2vec.

Note that by combining all layers together in this method, we have lost the details of each layer and also have no way to leverage the interactions between layers when the function $f$ is learned.

\begin{algorithm}[!t]
\caption{Network Aggregation Algorithm}
\label{Alg:NA}
\KwIn{Multilayer Network $MN$}
Initialize $G = (V,\emptyset)$\;
\For{all $i,j \in V$}
{
	\For{all $l\in L$}
    {
    \If{$a^l_{ij}=1$}{
        // add an edge for $G$\;
        $G$ $\leftarrow$ $G$ $\cup$ (i,j)\; 
        break}
    
    }
    
}

$f \leftarrow node2vec(G)$\;

\end{algorithm}

\subsection{Results Aggregation}
Here we project each network layer of the multilayer network into a separate vector space, and concatenate the resulting vector spaces. We define each layer graph as $G_l=(V,E_l)$ where $E_{ l }=\left\{ {(x,y)|a^l_{xy}=1} \right\}$ for $l\in L$ and learn $|L|$ functions $f_l: V\to \mathbb{R}^{d'}$ that are combined to form the map $f$ as $f=f_1||f_2||...||f_{|L|}$, where $||$ denotes the concatenation operator s.t. $f: V \to \mathbb{R}^{d'|L|}$.

Unlike the network aggregation method, the dimension $d=d'|L|$ of the resulting vector space scales with the number of layers in the layer set $L$. Though this method can in theory preserves the structure of each layer, it is also unable to leverage interactions between layers when learning the mapping function. Algorithm \ref{Alg:RA} outlines the result aggregation process.

\begin{algorithm}[!t]
\caption{Results Aggregation Algorithm}
\label{Alg:RA}

\KwIn{Multilayer Network $MN$}
initialize $f$ as empty\;
\For{ each $layer$ $l\in L$}
{
Initialize $G_l = (V,\emptyset)$\;
\For{all $i,j \in V$}
{

    \If{$a^l_{ij}=1$}{
        // add an edge for $G_l$\;
        $G_l$ $\leftarrow$ $G_l$ $\cup$ (i,j)\;
    }
    
    }
    $f_l\leftarrow node2vec(G_l)$\;
    $f\leftarrow f||f_l$
    
}


\end{algorithm}

\subsection{Layer Co-analysis}
To overcome the fact that neither network aggregation nor results aggregation can leverage interactions between layers, we adopt layer co-analysis for the construction of a vector space that is cognizant of interactions between layers as well as preserving the structure of each layer.

\begin{figure}[!t]
\centering

\subfigure[Path retention]{
	\label{subfig-path}
	\includegraphics[width=0.46\linewidth]{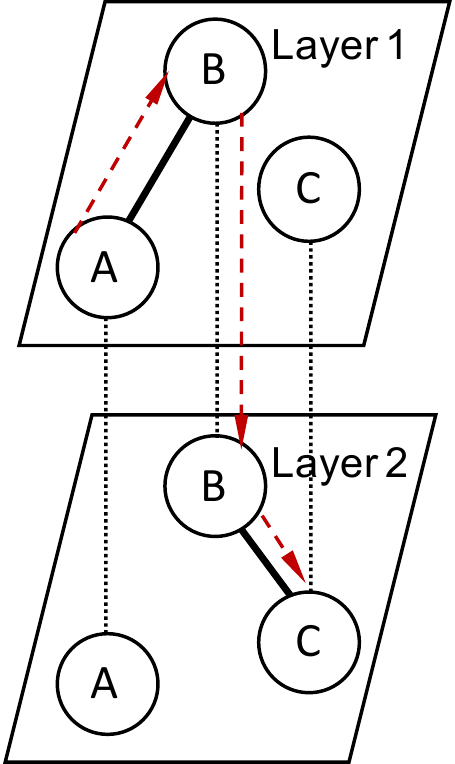}
}
\subfigure[Multi-edge retention]{
	\label{subfig-multiedge}
	\includegraphics[width=0.46\linewidth]{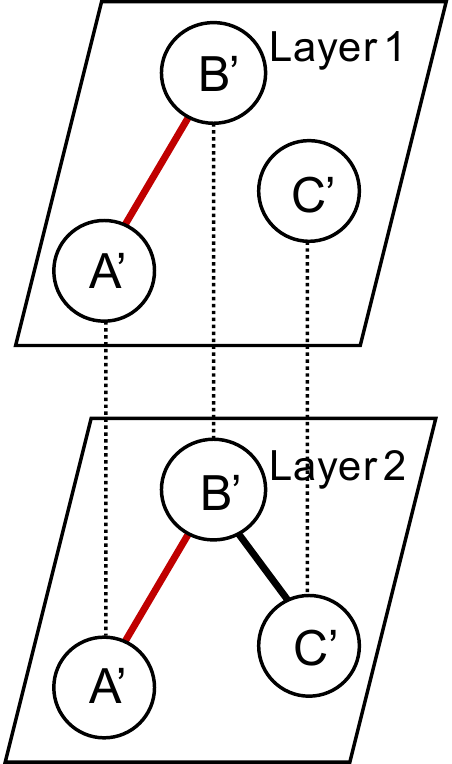}
}

\caption{Random walks on multilayer networks.}
\label{med-interplay}
\end{figure}

Consider random walks over the multilayer networks depicted in Figure \ref{med-interplay}, where dotted black lines represent correspondence between the nodes of each layer, and bold lines represent edges in the network. If we use result aggregation in Figure \ref{subfig-path}, there is no path from node $A$ to node $C$, as it is not possible to traverse between layers. Therefore the vector space cannot learn to associate these nodes. In contrast, if we run network aggregation on Figure \ref{subfig-multiedge}, we ignore the multi-edge between nodes $A'$ and $B'$ (represented by the bold red lines). These issues motivate the need for a method that can traverse the path (represented by the dotted red lines) between layers from $A$ to $C$, and that can retain the information implied by the multi-edge.


As LINE improved upon the uniform random walks of DeepWalk with weighted 2nd order walks, and node2vec introduced the parameters $p$ and $q$ to control the local and global biases of the sample random walks, we enable random walks to traverse between layers of a multilayer network, and introduce the parameter $r$ to control this tendency. 
Let $i_{|L|}$ represent the number of distinct layers connected to node $i$ $\forall i\in V$ of the multilayer network $MN$. That is, we can represent $i_{|L|}$ as in Equation \ref{eq:connected_layers}.
\begin{equation}
i_{ |L| }=\sum _{ l=1 }^{ |L| }{ I\left[ \left( \sum _{ (i,j,l)\in E }^{  }{ I[a^{ l }_{ ij }=1] }  \right) \geq 1 \right]  } 
\label{eq:connected_layers}
\end{equation}
Where $I$ denotes the indicator function. Then we introduce a 2nd order random walk with parameters $p,q,r$. If the random walk previously traversed edge $(z,x,l')$ and the current node $x$ only has edges on layer $l'$ (i.e. $x_{|L|}=1$), then we step according to the node2vec strategy (here, with binary edges), i.e. $P\left( t_{ i }=(x,y,l)|t_{ i-1 }={ (z,x,l') } \right)\propto\alpha_{pq}(z,x,l)$, where $\alpha_{pq}(z,x,l)$ is a multilayer modification to the node2vec $p$, $q$ factors as follows:
\begin{align}
\alpha _{ pq }(z,x,l)=\begin{cases} 1/p & if\quad d^l_{ zx}=0 \\ 1 & if\quad d^l_{ zx }=1 \\ 1/q & if\quad d^l_{zx}=2 \end{cases}
\end{align}
where $d^l_{zx}$ is the shortest path between nodes $z$ and $x$ in layer $l$ of the multilayer graph (where nodes $z,x$ may be the same node).

Otherwise, if $x_{|L|}>1$, the random walk stays on the current layer $l'$ with probability $r$, and moves along the edge of another layer $l$ with probability $1-r$. That is, the random walk traversal probability for $x_{|L|}>1$ is given by Equation \ref{eq:traversal}.

\begin{align}
&P\left( t_{ i }=(x,y,l)|t_{ i-1 }={ (z,x,l') } \right) \nonumber\\
\propto &\begin{cases} \alpha _{ pq }(z,x,l)r & if\quad l=l' \\ \frac { \alpha _{ pq }(z,x,l) }{ x_{ |L| }-1 } (1-r) & otherwise \end{cases}
\label{eq:traversal}
\end{align}

Pseudocode for this method is given in Algorithm \ref{Alg:Co}, where $node2vecSGD$ refers to running stochastic gradient descent on the node2vec loglikelihood \cite{grover2016node2vec} with the multilayer random walks taking the place of the standard node2vec walks. In this paper, we only consider binary edges (support for weighted multilayer networks could be incorporated trivially through multiplying the edge traversal probabilities by the normalized edge weights).

Note that in this algorithm, the $r$ variable represents how important we view the relationships between layers to be in comparison to the interactions between nodes of the same layer. 
For $r \rightarrow 0$, random walks will always traverse to different layers of the multilayer network when possible, whereas $r \rightarrow 1$ will restrict each random walk to stay on the layer in which it was initialized.

\begin{algorithm}[!t]
\caption{Layer Co-analysis Algorithm}
\label{Alg:Co}
\KwIn{Multilayer Network $MN$,$r,\alpha_{pq}$,$num\_walks$,$walk\_length$}
Initialize $walk\_list$ to empty\;
\For{$nw\_iter$ from 1 to $num\_walks$}
{
	Initialize current edge $(i,j,l)\leftarrow (i_0,j_0,l_0)$ uniformly at random\;
	\For{$wl\_iter$ from 1 to $walk\_length$}
    {
    $walk\_list[nw\_iter][wl\_iter]\leftarrow i$\;
    with probability $r$, choose $next\_layer=l$, otherwise choose $next\_layer=l'$ uniformly at random for some layer $l'$ incident to $j$\;
        set current edge $(i,j,l)\leftarrow(j,i',next\_layer)$ proportional to $\alpha_{pq}(j,i',next\_layer)$ for some $i'$ incident to $j$ through $next\_layer$\;

    }
}
$f\leftarrow node2vecSGD(walk\_list)$
\end{algorithm}




\section{Evaluation} \label{eva}
Here we choose five real-world multilayer datasets, comparing the performance of our three multilayer network embedding methods on the link prediction task. In this paper, we set $p=q=r=0.5$, with 10 random walks of 80 steps initialized from each node (i.e. we have $10|V|$ random walks in total for each graph), and compare against two standard methods of link prediction \cite{zhou2009predicting}. All experiments in this paper were conducted locally on CPU using a Mac Book Pro with an Intel Core i7 2.5GHz processor and 16GB of 1600MHz RAM.
Though this limits the size of our experiments in this preliminary work, our method naturally inherits the runtime scalability of node2vec.

\subsection{Datasets}
Table \ref{tab-datasets} shows the size of the five datasets, with layer information as follows:

\begin{table}[!t]
\centering
\caption{Information for five datasets.}
\scalebox{0.85}{
\begin{tabular}{|c|c|c|c|c|}
\hline
Datasets   & \# Layers & \# Nodes & \# Edges & \begin{tabular}[c]{@{}c@{}}labels (\# of corresponding nodes)\end{tabular}         \\ \hline
AUCS       & 5        & 61      & 353           & none                                                                                           \\ \hline
Terrorists & 4        & 78      & 623           & none                                                                                           \\ \hline
Students   & 3        & 185     & 311           & none                                                                                           \\ \hline
VC         & 3        & 29      & 250           & \begin{tabular}[c]{@{}c@{}}Boys (12)\\ Girls (17)\end{tabular}                                 \\ \hline
LN         & 4        & 191     & 511           & \begin{tabular}[c]{@{}c@{}}Leskovec's collaborator (87)\\ Ng's collaborator (104)\end{tabular} \\ \hline
\end{tabular}
}
\label{tab-datasets}
\end{table}

\begin{itemize}
\item AUCS \cite{kim2015community} (AUCS): The multiple layers represent five different relationship types
between 61 employees of a university department: (i) coworking, (ii) having
lunch together, (iii) Facebook friendship, (iv) onine
friendship (having fun together), and (v) coauthorship.

\item Terrorist network \cite{roberts2011roberts} (Terrorists): Each layer represents known interactions and ties between terrorists in the Noordin Top Terrorist dataset. These ties cover four different relationship types:
(i) communication, (ii) financial, (iii) operation, and (iv) trust.

\item Student cooperation \cite{fire2012predicting} (Students): Each layer represents a type of cooperation or coordination
between 185 students of Ben-Gurion University: (i) Computer Network, which represents students who finished their papers on the same machine. (ii) Partner's Network, which represents joint work on a submission.  (iii) Time Network, which indicates if students submitted papers in the same epoch.

\item Vickers Chan 7th grader social dataset \cite{vickers1981representing} (VC): Each layer represents an aspect of interaction between students in a class who were asked the following questions: (i) Who do you get on with in the class? (ii) Who are your best friends in the class? (iii) Who would you prefer to work with?

\item Leskovec-Ng collaboration dataset \cite{chen2017multilayer,zhang2014name,saha2015name}\footnote{The dataset can be downloaded from https://sites.google.com/site/pinyuchenpage/datasets} (LN): The coauthorship networks of Jure Leskovec and Andrew Ng from 1995 to 2014. A four layer multilayer graph is defined by partitioning the coauthorship networks into 5-year intervals. For each layer, there is an edge between two researchers if they coauthored at least one paper in the corresponding 5-year interval. In addition, each researcher is labeled as ``Leskovec’s collaborator'' or ``Ng’s collaborator'' depending upon collaboration frequency.
\end{itemize}

\subsection{Link Prediction Evaluation}
Here we perform the link prediction task with respect to the merged graph $G=(V,E)$ as defined in Section \ref{subsec:NA}. 
This is because the purpose of our proposed three methods is to leverage multilayer relationships of the datasets to inform the node embedding, this embedding does not inherently associate nodes (or edges between them) with particular layers of the multilayer network.

We split each edge set $E$ into a training subset $E^T$ and a test subset $E^{P}$, where $E=E^{T} \cup E^{P}, E^{T} \cap E^{P} = \emptyset$. In this paper, we randomly chose $10\%$ of the edges from each $E$ as $E^P$.

For a multilayer network trained on $E^{T}$, we use our proposed embedding methods to find the corresponding vector spaces. We then calculate the distances of node pairs corresponding to $E^{P}$, and reorder these distances into an ascending list. 
In order to predict links in the sampled multilayer network, we treat the first $10\%$ node pairs in $list$ have edges. Comparing these predicted edges with the true edges in $E^{P}$, we can evaluate the outcome of our methods.
Here, we introduce accuracy rate (see Equation \ref{eq-accuracy}) and F1-score (see Equation \ref{eq-f1}) to do the evaluation. 
First of all, accuracy rate indicates the number of edges $C$ that has been corrected estimated in $E^{P}$.
\begin{equation}
\label{eq-accuracy}
Accuracy = \frac{C}{|E^P|}
\end{equation}
Second, as F1-score is the harmonic mean of the precision and recall values for each layer, Equation \ref{eq-f1} evaluates the average F1-score of each layer. Here, $F-measure_{l}=\frac{2\cdot PREC_l \cdot RECALL_{l}}{PREC_l + RECALL_{l}}$, $PREC_l$ and $RECALL_l$ are the precision and recall values for each layer $l$. Larger F1-score means better prediction performance.
\begin{equation}
\label{eq-f1}
F_l(\{ C_l \}_{l=1}^{L}, \{ C'_l \}_{l=1}^{L'}) = \frac{1}{|L|} \sum F-measure_l
\end{equation}

In addition, we introduce two famous local link prediction methods \cite{zhou2009predicting} ``Common Neighbor Similarity'' and ``Jaccard Similarity'' on merged network as the comparison methods. Where Common Neighbor Similarity is defined in Equation \ref{eq-CN}, and Jaccard Similarity is defined in Equation \ref{eq-Ja}. Where $\Gamma(X)$, $\Gamma(Y)$ stands for the neighborhoods of the node $X,Y \in V$ in the merged graph.

\begin{equation}
\label{eq-CN}
CN_{xy} = |\Gamma(X) \cap \Gamma(Y)|
\end{equation}
\begin{equation}
\label{eq-Ja}
Jaccard_{xy} = \frac{|\Gamma(X) \cap \Gamma(Y)|}{|\Gamma(X) \cup  \Gamma(Y)|}
\end{equation}

Table \ref{tab-lp} shows the accuracy and F1-Score results for different datasets. In addition, we use bold text to indicate the best performance for each datasets. From the table we can tell that except LN datasets, our methods can achieve higher accuracy and F1-score for the rest of the datasets.

\begin{figure*}[!t]
\centering
\includegraphics[width=\linewidth]{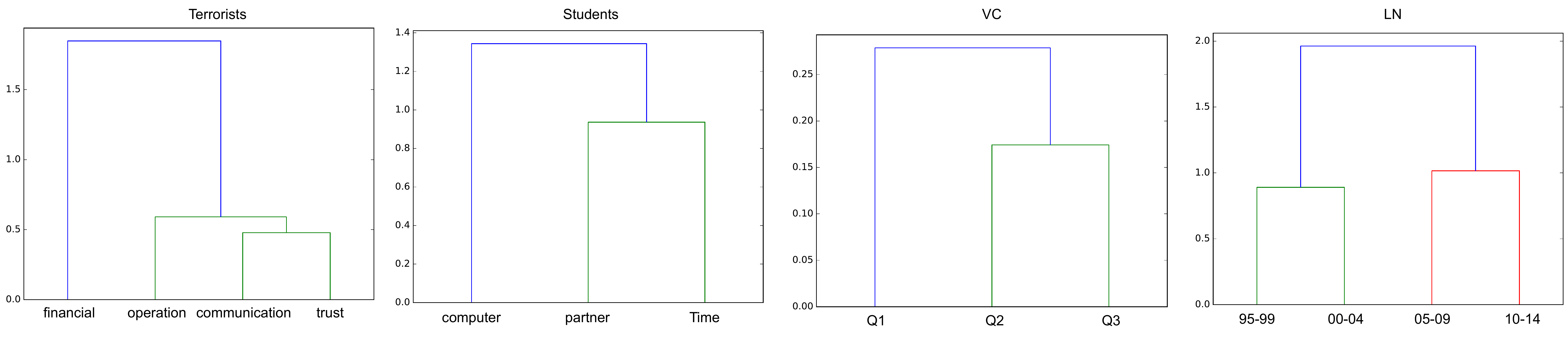}
\caption{Layers Distance for the four datasets.}
\label{eva-layer}
\end{figure*}

\begin{figure*}[!t]
\centering
\subfigure[Four Layers of Terrorists Dataset]{
	\label{eva-terrorist-graphs}	
    \includegraphics[width=\linewidth]{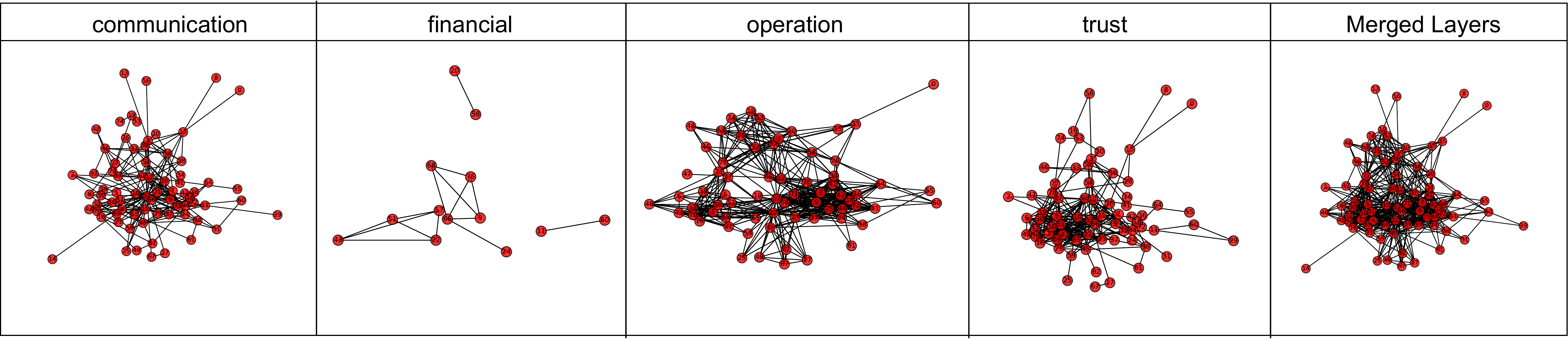}
}
\subfigure[Three Layers of VC Dataset]{
	\label{eva-VC-graphs}	
    \includegraphics[width=\linewidth]{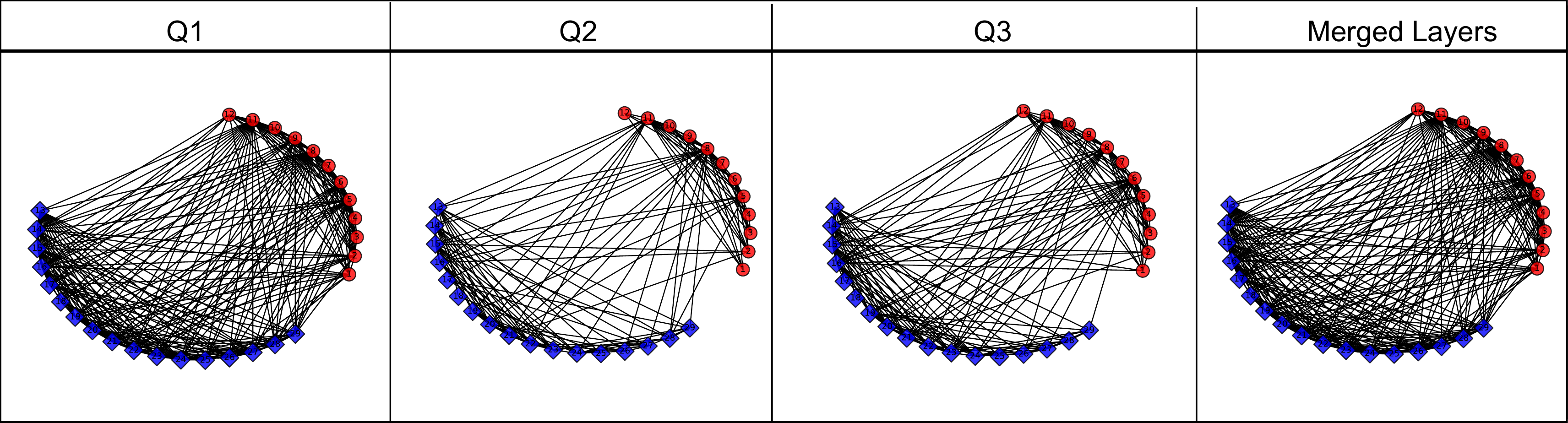}
}
\subfigure[Four Layers of LN Dataset]{
	\label{eva-LN-graphs}	
    \includegraphics[width=\linewidth]{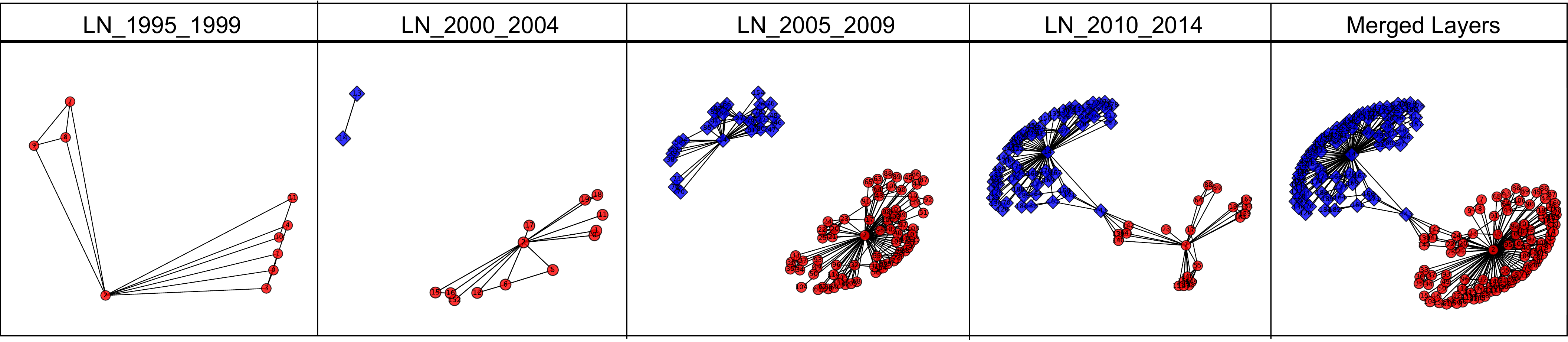}
}
\caption{Topology of three multilayer networks.}
\label{eva-topology}
\end{figure*}

We use Figure \ref{intro-layers} and Figure \ref{eva-layer} to show the layer distance for each corresponding dataset.
In addition, we use Figure \ref{eva-topology} to show topology of each layer and the corresponding merged layers of three multilayer networks. As VC and LN have label information, we use different shape and color to demonstrate nodes with different labels. 
Combine all these evaluations, we give a detailed analysis for each datasets.

\begin{itemize}
\item For AUCS, Terrorists and Students datasets: As AUCS dataset represents the interactions activities among employees, Terrorists dataset shows how terrorists work together, and Students dataset indicates the collaboration among students. It is clear that different layers have strong influence with each other. Which means if we want to predict edges information among two nodes, the important interactions of these two nodes among different layers should be considered. What's more, take the topology of layers of Terrorists dataset in Figure \ref{eva-terrorist-graphs} as an example,  there are strong interactions among four layers, although layer 2 (Financial) has a small number of nodes, but layer co-analysis can recover necessarily information by random walk on different layers. Hence, it is reasonable and important to consider the interactions for different layers.

\item For VC dataset: As different questions indicate different problems, these different layers have relatively weak connections. For example, we cannot argue that a person gets on with (Q1) are all of his/her best friends, while the best friends of a person (Q2) are the same group of people that the person wants to work with (Q3). As shown in Figure \ref{eva-VC-graphs}, the first question is too general, which causes to create lots of noises (unnecessary edges) and therefore cannot indicate the true relationships among these nodes. If we combine these layers together or put more concentration on interactions among layers, then neither network aggregation nor layer co-analysis can reveal true information instead of just introducing more noises into the analysis.

\item For LN dataset: this dataset is a temporal dataset across 20 years that people joins or leaves the Leskovec's group or Andrew Ng's group. As shown in Figure \ref{eva-LN-graphs}, different layers in different time do not show any interactions. For example, in the first layer (LN\_1995\_1999), there is no blue nodes, and for the second (LN\_2000\_2004) and third layer (LN\_2005\_2010), two groups are expanded by themselves. what's more, as the time span of the multilayer network is too large, this particular feature indicates the fact that the interaction among these layers is not the key reason to form the topology in each corresponding layer. What's more, as there are a 5 years span between layers, the noises in these layers are the major reason why our methods cannot function well. Instead, the original Jaccard method is the best. Because this method only cares about the average number of shared neighbors for two nodes. So the noise has the least affected for such method.
\end{itemize}

\begin{table*}[!t]
\centering
\caption{Accuracy and F1-Score for Different Methods}
\begin{tabular}{|c|c|c||c|c|c|}
\hline
\multirow{3}{*}{Datasets} & \multicolumn{5}{c|}{Accuracy / F1-Score for Different Methods}                                          \\ \cline{2-6} 
						  & \multicolumn{2}{c||}{Regular Link Prediction Methods} 
                          & \multicolumn{3}{c|}{Our Methods}        \\ \cline{2-6} 
                          & Common Neighbor & Jaccard Similarity & Network Aggregation & Results Aggregation & Networks Co-analysis \\ \hline
AUCS                      & 0.029 / 0.056   & 0.051 / 0.097      & 0.184 / 0.311       & 0.092 / 0.168       & \textbf{0.207} / \textbf{0.343}        \\ \hline
Terrorists                & 0.012 / 0.024   & 0.016 / 0.032      & 0.229 / 0.373       & 0.090 / 0.166       & \textbf{0.347} / \textbf{0.515}        \\ \hline
Students                  & 0.015 / 0.030   & 0.138 / 0.243      & 0.139 / 0.225      & 0.063 / 0.119       & \textbf{0.127} / \textbf{0.2444}        \\ \hline
VC                        & 0.049 / 0.093   & 0.098 / 0.178      & 0.400 / 0.571       & \textbf{0.650} / \textbf{0.788}       & 0.550 / 0.710        \\ \hline
LN                        & 0.027 / 0.053   & \textbf{0.206} / \textbf{0.342}      & 0.103 / 0.187       & 0.070 / 0.130       & 0.083 / 0.153        \\ \hline
\end{tabular}
\label{tab-lp}
\end{table*}

\section{Related Work on Random Walk based Network Embedding}\label{sec:related}
In this section, we review related work on standard network embedding, that is, embedding methods proposed for single-layer graphs.
With the development of unsupervised feature learning techniques \cite{bengio2013representation}, deep learning methods proved successful in natural language processing tasks through neural language models. These models have been used to capture the semantic and syntactic structures of human language \cite{collobert2008unified}, and even logical analogies \cite{mikolov2013linguistic}, by embedding words as vectors.
As a graph can be interpreted as a kind of language (by treating random walks as the equivalent of sentences),
DeepWalk \cite{perozzi2014deepwalk} introduced such methods into network analysis, allowing for the projection of network nodes into a vector space. To solve the scalability problem of this method when applied to real world information networks (which often contain millions of nodes), LINE \cite{tang2015line} was developed. LINE extended the uniform random walks of DeepWalk to 1st and 2nd order weighted random walks, and it can project a network with millions of vertices and billions of edges into a vector space in a few hours.

However, both methods have limitations.
As DeepWalk uses uniform random walks for searching, it cannot provide control over the explored neighborhoods. In contrast, LINE proposes a breadth-first strategy to sample nodes and optimize the likelihood independently over 1-hop and 2-hop neighbors, but it has no flexibility in exploring nodes at future depths. In order to deal with both of these limitations, node2vec \cite{grover2016node2vec} provides a flexible and controllable strategy for exploring network neighborhoods through the parameters $p$ and $q$. From a practical standpoint, node2vec is scalable and robust to perturbations. Of course, none of these methods can deal with random walk samples that intelligently consider traversals between layers of multilayer networks. One of our methods (layer co-analysis) is therefore a natural progression of the literature in extending the capabilities of the random walk samples.

\section{Discussion and Future Work}\label{sec:conclusion}
In this paper, we demonstrate three different methods to project multilayer network into a representative vector space. The first method (Network Aggregation) aggregates all layers together to construct a merged network, and use standard network embedding method to project a multilayer network into a vector space.
The second method (Results Aggregation) uses standard network embedding to obtain a vector space for each corresponding layer, and then combines these vector spaces together to construct a new vector space for the multilayer network.
At last, as the first two methods do not leverage the important interactions between layers, we introduce a layer co-analysis method which leverage interactions among layers. In layer co-analysis, we use $r$ to constrain the behavior of the walk, where the greater the $r$, the greater the chance of the random walk to stay in the same layer. On the contrary, the smaller the $r$, the greater the possibility of random walk to choose different layers.
In the evaluation part, we compare the accuracy and F1-score for five datasets, by comparing to regular link prediction methods, we have proved that our method do have the ability to project a multilayer network into the suitable vector space.

To the best of our knowledge, since this paper is a first-line research for principled graph embedding a multilayer network into a vector space, our experimental results suggest some future work and new challenges along this line:
(i) From evaluation aspect, as we only use link prediction as the evaluation in this paper, the performance on multi-label classification is worth exploring. In addition, as our methods can be simply applied to layers with weighted edges and weighted interactions, we will test the performance on weighted multilayer network.
(ii) From algorithm perspective, the proposed co-layer analysis method involves an additional layer transition probability $r$ for multilayer network embedding. In the future work, we will further discuss how to automatically learn $r$ by analyzing layer distance for a multilayer network.
(iii) From the data type perspective, as attributed graphs have been widely introduced in big data analysis, we will continue to discuss the possibility to project an attributed multilayer graphs into a proper vector space by taking into the node/edge's properties.






%


\newpage
\bibliographystyle{IEEEtran}
\bibliography{paper.bib} 

\begin{thebibliography}{10}
\providecommand{\url}[1]{#1}
\csname url@samestyle\endcsname
\providecommand{\newblock}{\relax}
\providecommand{\bibinfo}[2]{#2}
\providecommand{\BIBentrySTDinterwordspacing}{\spaceskip=0pt\relax}
\providecommand{\BIBentryALTinterwordstretchfactor}{4}
\providecommand{\BIBentryALTinterwordspacing}{\spaceskip=\fontdimen2\font plus
\BIBentryALTinterwordstretchfactor\fontdimen3\font minus
  \fontdimen4\font\relax}
\providecommand{\BIBforeignlanguage}[2]{{%
\expandafter\ifx\csname l@#1\endcsname\relax
\typeout{** WARNING: IEEEtran.bst: No hyphenation pattern has been}%
\typeout{** loaded for the language `#1'. Using the pattern for}%
\typeout{** the default language instead.}%
\else
\language=\csname l@#1\endcsname
\fi
#2}}
\providecommand{\BIBdecl}{\relax}
\BIBdecl

\bibitem{boccaletti2014structure}
S.~Boccaletti, G.~Bianconi, R.~Criado, C.~I. Del~Genio, J.~G{\'o}mez-Gardenes,
  M.~Romance, I.~Sendina-Nadal, Z.~Wang, and M.~Zanin, ``The structure and
  dynamics of multilayer networks,'' \emph{Physics Reports}, vol. 544, no.~1,
  pp. 1--122, 2014.

\bibitem{de2013mathematical}
M.~De~Domenico, A.~Sol{\'e}-Ribalta, E.~Cozzo, M.~Kivel{\"a}, Y.~Moreno, M.~A.
  Porter, S.~G{\'o}mez, and A.~Arenas, ``Mathematical formulation of multilayer
  networks,'' \emph{Physical Review X}, vol.~3, no.~4, p. 041022, 2013.

\bibitem{loe2015comparison}
C.~W. Loe and H.~J. Jensen, ``Comparison of communities detection algorithms
  for multiplex,'' \emph{Physica A: Statistical Mechanics and its
  Applications}, vol. 431, pp. 29--45, 2015.

\bibitem{kim2015community}
J.~Kim and J.-G. Lee, ``Community detection in multi-layer graphs: A survey,''
  \emph{ACM SIGMOD Record}, vol.~44, no.~3, pp. 37--48, 2015.

\bibitem{de2015structural}
M.~De~Domenico, V.~Nicosia, A.~Arenas, and V.~Latora, ``Structural reducibility
  of multilayer networks,'' \emph{Nature communications}, vol.~6, 2015.

\bibitem{de2017community}
C.~De~Bacco, E.~A. Power, D.~B. Larremore, and C.~Moore, ``Community detection,
  link prediction and layer interdependence in multilayer networks,''
  \emph{arXiv preprint arXiv:1701.01369}, 2017.

\bibitem{heaney2014multiplex}
M.~T. Heaney, ``Multiplex networks and interest group influence reputation: An
  exponential random graph model,'' \emph{Social Networks}, vol.~36, pp.
  66--81, 2014.

\bibitem{hu2014conditions}
Y.~Hu, S.~Havlin, and H.~A. Makse, ``Conditions for viral influence spreading
  through multiplex correlated social networks,'' \emph{Physical Review X},
  vol.~4, no.~2, p. 021031, 2014.

\bibitem{costenbader2003stability}
E.~Costenbader and T.~W. Valente, ``The stability of centrality measures when
  networks are sampled,'' \emph{Social networks}, vol.~25, no.~4, pp. 283--307,
  2003.

\bibitem{brodka2011degree}
P.~Br{\'o}dka, K.~Skibicki, P.~Kazienko, and K.~Musia{\l}, ``A degree
  centrality in multi-layered social network,'' in \emph{Computational Aspects
  of Social Networks (CASoN), 2011 International Conference on}.\hskip 1em plus
  0.5em minus 0.4em\relax IEEE, 2011, pp. 237--242.

\bibitem{brodka2012analysis}
P.~Br{\'o}dka, P.~Kazienko, K.~Musia{\l}, and K.~Skibicki, ``Analysis of
  neighbourhoods in multi-layered dynamic social networks,''
  \emph{International Journal of Computational Intelligence Systems}, vol.~5,
  no.~3, pp. 582--596, 2012.

\bibitem{kazienko2010individual}
P.~Kazienko, P.~Brodka, and K.~Musial, ``Individual neighbourhood exploration
  in complex multi-layered social network,'' in \emph{Web Intelligence and
  Intelligent Agent Technology (WI-IAT), 2010 IEEE/WIC/ACM International
  Conference on}, vol.~3.\hskip 1em plus 0.5em minus 0.4em\relax IEEE, 2010,
  pp. 5--8.

\bibitem{brodka2010method}
P.~Br{\'o}dka, K.~Musial, and P.~Kazienko, ``A method for group extraction in
  complex social networks,'' \emph{Knowledge Management, Information Systems,
  E-Learning, and Sustainability Research}, pp. 238--247, 2010.

\bibitem{zhou2005maximal}
T.~Zhou, G.~Yan, and B.-H. Wang, ``Maximal planar networks with large
  clustering coefficient and power-law degree distribution,'' \emph{Physical
  Review E}, vol.~71, no.~4, p. 046141, 2005.

\bibitem{de2015identifying}
M.~De~Domenico, A.~Lancichinetti, A.~Arenas, and M.~Rosvall, ``Identifying
  modular flows on multilayer networks reveals highly overlapping organization
  in interconnected systems,'' \emph{Physical Review X}, vol.~5, no.~1, p.
  011027, 2015.

\bibitem{domenico2015identifying}
M.~Domenico, A.~Lancichinetti, A.~Arenas, and M.~Rosvall, ``Identifying modular
  flows on multilayer networks reveals highly overlapping organization in
  social systems,'' \emph{Phys. Rev}, vol.~5, p. 011027, 2015.

\bibitem{chen2016multilayer}
P.-Y. Chen and A.~O. Hero~III, ``Multilayer spectral graph clustering via
  convex layer aggregation,'' \emph{arXiv preprint arXiv:1609.07200}, 2016.

\bibitem{jeub2017local}
L.~G. Jeub, M.~W. Mahoney, P.~J. Mucha, and M.~A. Porter, ``A local perspective
  on community structure in multilayer networks,'' \emph{Network Science}, pp.
  1--20, 2017.

\bibitem{deford2017spectral}
D.~R. DeFord and S.~D. Pauls, ``Spectral clustering methods for multiplex
  networks,'' \emph{arXiv preprint arXiv:1703.05355}, 2017.

\bibitem{benson2016higher}
A.~R. Benson, D.~F. Gleich, and J.~Leskovec, ``Higher-order organization of
  complex networks,'' \emph{Science}, vol. 353, no. 6295, pp. 163--166, 2016.

\bibitem{boguna2009navigability}
M.~Boguna, D.~Krioukov, and K.~C. Claffy, ``Navigability of complex networks,''
  \emph{Nature Physics}, vol.~5, no.~1, pp. 74--80, 2009.

\bibitem{chen2017fast}
S.~Chen, S.~Niu, L.~Akoglu, J.~Kova{\v{c}}evi{\'c}, and C.~Faloutsos, ``Fast,
  warped graph embedding: Unifying framework and one-click algorithm,''
  \emph{arXiv preprint arXiv:1702.05764}, 2017.

\bibitem{perozzi2014deepwalk}
B.~Perozzi, R.~Al-Rfou, and S.~Skiena, ``Deepwalk: Online learning of social
  representations,'' in \emph{Proceedings of the 20th ACM SIGKDD international
  conference on Knowledge discovery and data mining}.\hskip 1em plus 0.5em
  minus 0.4em\relax ACM, 2014, pp. 701--710.

\bibitem{tang2015line}
J.~Tang, M.~Qu, M.~Wang, M.~Zhang, J.~Yan, and Q.~Mei, ``Line: Large-scale
  information network embedding,'' in \emph{Proceedings of the 24th
  International Conference on World Wide Web}.\hskip 1em plus 0.5em minus
  0.4em\relax ACM, 2015, pp. 1067--1077.

\bibitem{grover2016node2vec}
A.~Grover and J.~Leskovec, ``node2vec: Scalable feature learning for
  networks,'' in \emph{Proceedings of the 22nd ACM SIGKDD International
  Conference on Knowledge Discovery and Data Mining}.\hskip 1em plus 0.5em
  minus 0.4em\relax ACM, 2016, pp. 855--864.

\bibitem{berlingerio2013abacus}
M.~Berlingerio, F.~Pinelli, and F.~Calabrese, ``Abacus: frequent pattern
  mining-based community discovery in multidimensional networks,'' \emph{Data
  Mining and Knowledge Discovery}, vol.~27, no.~3, pp. 294--320, 2013.

\bibitem{zhou2009predicting}
T.~Zhou, L.~L{\"u}, and Y.-C. Zhang, ``Predicting missing links via local
  information,'' \emph{The European Physical Journal B-Condensed Matter and
  Complex Systems}, vol.~71, no.~4, pp. 623--630, 2009.

\bibitem{roberts2011roberts}
N.~Roberts, ``Roberts and everton terrorist data: Noordin top terrorist network
  (subset),'' \emph{Machinereadable data file}, 2011.

\bibitem{fire2012predicting}
M.~Fire, G.~Katz, Y.~Elovici, B.~Shapira, and L.~Rokach, ``Predicting student
  exam’s scores by analyzing social network data,'' in \emph{International
  Conference on Active Media Technology}.\hskip 1em plus 0.5em minus
  0.4em\relax Springer, 2012, pp. 584--595.

\bibitem{vickers1981representing}
M.~Vickers and S.~Chan, ``Representing classroom social structure,''
  \emph{Victoria Institute of Secondary Education, Melbourne}, 1981.

\bibitem{chen2017multilayer}
P.-Y. Chen and A.~O. Hero, ``Multilayer spectral graph clustering via convex
  layer aggregation: Theory and algorithms,'' \emph{IEEE Transactions on Signal
  and Information Processing over Networks}, 2017.

\bibitem{zhang2014name}
B.~Zhang, T.~K. Saha, and M.~Al~Hasan, ``Name disambiguation from link data in
  a collaboration graph,'' in \emph{Advances in Social Networks Analysis and
  Mining (ASONAM), 2014 IEEE/ACM International Conference on}.\hskip 1em plus
  0.5em minus 0.4em\relax IEEE, 2014, pp. 81--84.

\bibitem{saha2015name}
T.~K. Saha, B.~Zhang, and M.~Al~Hasan, ``Name disambiguation from link data in
  a collaboration graph using temporal and topological features,'' \emph{Social
  Network Analysis and Mining}, vol.~5, no.~1, p.~11, 2015.

\bibitem{bengio2013representation}
Y.~Bengio, A.~Courville, and P.~Vincent, ``Representation learning: A review
  and new perspectives,'' \emph{IEEE transactions on pattern analysis and
  machine intelligence}, vol.~35, no.~8, pp. 1798--1828, 2013.

\bibitem{collobert2008unified}
R.~Collobert and J.~Weston, ``A unified architecture for natural language
  processing: Deep neural networks with multitask learning,'' in
  \emph{Proceedings of the 25th international conference on Machine
  learning}.\hskip 1em plus 0.5em minus 0.4em\relax ACM, 2008, pp. 160--167.

\bibitem{mikolov2013linguistic}
T.~Mikolov, W.-t. Yih, and G.~Zweig, ``Linguistic regularities in continuous
  space word representations.'' in \emph{Hlt-naacl}, vol.~13, 2013, pp.
  746--751.

\end{thebibliography}

\end{document}